Virus evolution : from neglect to center stage, amidst some confusion

Jean-Michel Claverie

Structural & Genomic Information Laboratory
CNRS-UPR2589, IBSM
Parc Scientifique de Luminy
163 Avenue de Luminy, case 934
Marseille 13288, cedex 9 , France

E-mail : Jean-Michel.Claverie@igs.cnrs-mrs.fr

Reputed intractable, the question of the origin of viruses has long been neglected. In the modern literature "Virus evolution" has come to refer to study more akin to population genetics, such as the world-wide scrutiny on new polymorphisms appearing daily in the H5N1 avian flu virus [1], than to the fundamental interrogation: where do viruses come from? This situation is now rapidly changing, due to the coincidence of bold new ideas (and sometimes the revival of old ones), the unexpected features exhibited by recently isolated spectacular viruses [2] (see at URL: www.giantvirus.org), as well as the steady increase of genomic sequences for "regular" viruses and cellular organisms enhancing the power of comparative genomics [3]. After being considered non-living and relegated in the wings by a majority of biologists, viruses are now pushed back on the center stage: they might have been at the origin of DNA, of the eukaryotic cell, and even of today's partition of biological organisms into 3 domains of life: bacteria, archaea and eukarya. Here, I quickly survey some of the recent discoveries and the new evolutionary thoughts they have prompted, before adding to the confusion with one interrogation of my own: what if we totally missed the true nature of (at least some) viruses?

As of April 2006, more than 1600 viral genomes have been sequenced, approximately equally divided into RNA and DNA viruses. In view of this fundamental difference about the nature of their genetic material (hence replication mechanisms, but also size, genetic complexity, host range,…, etc.) it is quite tempting to immediately rule out the idea that they derive from a common ancestor (i.e. be "monophyletic"). However that might not be so simple. Although there are many arguments in favor of these viruses to have been invented separately, (RNA viruses first, in the context of the "RNA world" theory), their genesis might have overlapped quite significantly before or shortly after the Last Universal Common Ancestor cell (the last unique ancestor of all living beings) [reviewed in 2], allowing a non negligible level of genome mixing. Indeed, several proteins have homologues found both in RNA and DNA viruses, the most important of all being the jelly-roll capsid protein [4], the sole component uniquely distinguishing the viruses from the cellular world [5]. Other unexpected -albeit more sporadic- sharing of homologous components between the two types of viruses are frequent, however, they are considered the results of more recent lateral gene transfers, such as the chaperonin Hsp70 uniquely found in giant ds-DNA Mimivirus [6] and positive-strand RNA closteroviruses [7].

The notion that viruses might be very ancient (and even ancestral to cells as proposed by d'Herelle, the discoverer of bacteriophages [8]), once modernized to take into account all our present knowledge of molecular biology and genomics [9, 10], has become the starting point of increasingly daring evolutionary scenarios. To explain the puzzling phylogenies and distribution of many DNA informational proteins between the three domains of life, it has now been proposed that DNA viruses could be at the origin of today's eukaryotic replication proteins [11, 12]. At about the same time, others postulated that a large poxvirus-like ds-DNA virus might be at the origin of the whole eukaryotic nucleus, as an organelle [13, 14]. I personally found the global idea that, inside a cell, a nucleus is functionally equivalent to a selfish DNA virus (i.e. replicating "its" DNA using the cell metabolism) simple and very appealing. Even more so when one realizes that it can be used in a reversible way: the nucleus of a (primitive) eukaryote (re-) turning into a large DNA virus (Fig.1). Most interestingly, such transfer of "infectious" nucleus is a well documented phenomenon in many parasitic red algal species [15]. Such back-and-forth eukaryogeneis-viriogenesis scenario could readily explain the multiplicity of today's virus lineages, together with their diversity in size, complexity, and gene complement, as well as the apparent mixture of monophyly and polyphyly exhibited by the viral world. In this context, extant complex DNA eukaryotic viruses could have originated from iterative waves of "nuclear viriogenesis". However, we still need some initial "seeding" virus, the one that, for instance, invented the prototype of the now nearly ubiquitous jelly-roll capsid protein. Reviving the initial d'Herelle "virus first" hypothesis, Koonin and Martin [16] paradoxically proposed that RNA viruses might have emerged even prior to the invention of individual cells, as a selfish RNA replicons roaming prebiotic inorganic compartments. However, there is little chance that this hypothesis could be scientifically proven anytime soon.

Also quite provocative, is the idea that RNA viruses might be at the origin of DNA biochemistry [2, 17]. According to this scenario, RNA-based viruses infecting RNA-based cell would have acquired a RNA to DNA modification system to resist cellular RNA degrading enzymes (e.g. the RNA equivalent of today's restriction/modification systems). For this purpose, RNA viruses would thus have invented ribonucleotide reductase (making deoxyribo-XDP from ribo-XDP) and thymidylate synthase (making dTMP from dUMP) the two key-pathways in DNA synthesis. Cellular RNA was then replaced by DNA in the course of evolution because of its better stability, and the repairing capability of its double-stranded structure, allowing larger, more complex genomes to compete out the more primitive RNA cells. Notice that this scenario is nicely complementary to the above viral eukaryogenesis hypothesis, the cellular RNA genes being progressively recruited within the DNA-based newly acquired "nucleus" (Fig. 1). Interestingly, deoxyuridylic acid has been shown to replace thymidylic acid in the DNA of several bacteriophages [18].

Lastly, in a paper that already received much attention, Forterre promoted (ancient) viruses to another fundamental role: to have been at the origin of the three basic cellular domains [19]. His "three RNA cells, three DNA viruses" hypothesis, explains 1) why there are three discrete lineages of modern cells instead of a continuum, 2) the existence of three canonical ribosomal patterns, and 3) the critical differences exhibited by the nevertheless similar eukaryal and archaea replication machineries. This is readily done by postulating that DNA technology was independently transferred by three different founder DNA viruses to RNA-based Archaea, Bacteria, and Eukarya ancestors respectively. The reduction of evolutionary rates following the transition from RNA to DNA genomes would have respectively stabilized the three canonical versions of proteins involved in translation that are still highly recognizable today.

If, for a moment, we put aside the paradoxical "virus-first" hypothesis, we are left with two more traditional ("cell-first") hypotheses about the origin of viruses in general: i) the *escape hypothesis*, viewing viruses as originating from the escape of the minimal set of cellular components necessary to constitute an infectious replicating selfish system, and ii) the *reduction hypothesis*, in which they would have derived from a cellular organism through a progressive loss of functions until becoming a *bona fide* virus. In real life, unfortunately, this simple dichotomy will be blurred by the accretion of laterally transferred genes between virus (or parasitic cellular organisms) sharing identical hosts, or directly captured from the virus hosts. In that respect, bacteriophages markedly differs from most eukaryotic ds-DNA virus by exhibiting massive recombinational reassortments and accretion of genes, most likely resulting from the existence of a host integrated prophage state [20]. Yet, 80% of the ds-DNA bacteriophage genes still have no obvious homologues in microbial genomes, suggesting a large degree of evolutionary independence of the phage gene set [21]. A much stricter genetic isolation is exhibited by the eukaryotic nucleo-cytoplasmic large ds-DNA virus (NCLDV), such as the giant *Acanthamoeba polyphaga* mimivirus [6], the 1.2 Mb genome (911 genes) of which exhibits little evidence of horizontal transfer [22]. This also holds true for the next largest NCLDVs, algae-infecting phycodnaviruses (with known genome sequences in the 300-400kb range) [23, 24]. Mimivirus also exhibits a high level of genomic coherence, as shown by the homogeneity of its nucleotide composition and the strict conservation of half of its promoter sequences [25].

As more genomes of large eukaryotic viruses are sequenced, new genes keep turning out, most of them with no obvious phylogenetic affinity with known hosts or extant cellular organisms. This simple observation is definitely more in favor of these large viruses arising from the reduction of a more complex ancestral (viral) genome, than supporting the hypothetical accretion of numerous exogenous genes (without recognizable origin) around a primitive minimal viral genome. Recent results on Coccolithovirus EhV-86 illustrate this point very nicely. Until the 407-kb genome of EhV-86 was characterized, the trademark of all previously characterized phycodnaviruses (with smaller 320-kb genomes), compared to other NCLDVs was the absence of a virus-encoded transcription machinery (lack of DNA-directed RNA polymerase) [23]. Obviously, the presence/absence of a RNA polymerase implies major differences in virus physiology. Unexpectedly, EhV-86 was found to encode its own 6-subunit transcriptional machinery [24]. Yet, a phylogenetic analysis of 25 core genes common to NCLDVs firmly placed this EhV-86 within the Phycodnaviridae clade [24]. In this case, the loss of the transcription apparatus by the smaller phycodnaviruses, rather than the simultaneous gain of the 6-subunit of a RNA polymerase by EhV-86, appears much more likely.

The reduction hypothesis received a strong boost from the discovery and genome characterization of *Acanthamoeba polyphaga* mimivirus [6], the first virus to largely overlap with the world of cellular organisms, both in term of particle size and genome complexity [2]. The finding of numerous virally encoded components of a nevertheless largely incomplete translation apparatus, strongly suggested a process of reductive evolution, from an even more complex ancestor, eventually endowed with protein synthesis capacity. Such an ancestor, either could have evolved from an obligate intracellular parasitic cell (functionally similar to a Rickettsia or a Chlamydia), or be derived from the nucleus of a primitive eukaryote through the mechanism illustrated in Fig.1. If the reduction scenario is at the origin of Mimivirus, it most likely applies to other NCLDVs, in particular to those exhibiting the best phylogenetic affinity such as the Phycodnaviridae and Iridoviridae. Sequencing additional large genomes from representatives of these families should provide valuable insights about this postulated giant ancestor.

At first sight, bacterial intracellular obligate parasites such as Rickettsia or Chlamydia have little functional resemblance with Mimivirus despite exhibiting a comparable genomic complexity. On one side, one has metabolically active organisms, stealing ATP and biochemical precursor from their hosts, to transcribe their genomes, translate their proteins, replicate their DNA, and divide. On the other side, one sees a large but metabolically silent viral particle, not deserving to be qualified as "living" by a majority of biologists. However, this traditional view might be a case of "when the finger points to the stars, the fool looks at the finger". Rather than comparing a parasitic cell to the virus *particle*, we should compare it to the virus *factory* [26]. Not much is yet known specifically on Mimivirus factories, but upon infection, all complex eukaryotic viruses such as iridoviridae, poxviridae, and asfarviridae, give raise to complex intracellular structures that transcribe the viral genome, translate transcripts into proteins, replicate the viral DNA, before packaging it into sophisticated vehicles designed to reproduce the virus factory upon infection of another host cell (Fig. 2). The virus factory is enclosed by a membrane (often derived from the rough endoplasmic reticulum) to exclude cellular organelles, but contains ribosomes and cytoskeletal elements. In the meantime, virus factory recruits mitochondria at their periphery, from which they obtain ATP [26]. The overall functional resemblance between an intracellular parasitic bacteria and a large eukaryotic virus *at this stage*, is quite striking. From this point of view, the genomic complexity of NCLDVs is no longer paradoxical, as it is commensurate to the complexity of the cell-like virus factory, but not to the particle used to reproduce it. Interpreting the virion particle as the "virus", is very much like looking at a spermatozoid and calling it a "human": clearly a 3,000-Mb genome would seem like an overkill for such a unicellular organism (as for plant genomes when looking at metabolically inert pollen grains).

Conceptually, the analogy between a virus life cycle and the reproductive cycle of a non-dividing organism can be extended farther. *Sensu* August Weismann, the virus particle possesses all the property of the Germen (the continuous immortal lineage responsible for carrying one generation to the next) while the transient virus factory exhibits all the property of the Soma [27]. Also, according to Weismann, such a partition implies the phenomenon of aging: once the opportunity to pass germplasm on has passed (i.e. viral particles were produced), there is no need to maintain the integrity of the somaplasm. With this interpretation, the virus factory now becomes the ultimate illustration of a disposable soma, vanishing right after viral particles have been produced. Nevertheless, it should be considered the "actual virus organism" when referring to a "virus". Incidentally, this interpretation makes the living nature of viruses undisputable, on the same footing than bacterial intracellular parasites. Focusing on the structure of the virus factory rather than on the morphology of the virus particle might help reach a better understanding of the evolutionary history of viruses.

A serious difficulty in the reductive hypothesis for the origin of viruses (seen as particles) is to propose reasonable mechanisms by which a cell, even highly parasitic, might switch from a cellular-dividing mode to a host-supported particle-replication mode, at once. Focusing on viruses as cell-like factory rather than as particles, makes it much easier to conceive a gradual transition. The event committing a parasitic cell toward the viral reductive evolution pathway would be the loss of an essential component (e.g. a ribosomal protein) of its translation apparatus: the presence/absence of an encoded protein synthesis system clearly remains the last unambiguous genomic divide between the viral and the cellular worlds. In order to survive, the now translation-defective cell had to adopt new strategies to gain access to the ribosomes of its hosts. At the same time, this translation-defective cell could now dispense with the rest of its ribosome encoding genes. Such intermediary protoviral cell could survive in its original host, while improving the design of a *bona fide* virus factory. Finally, a gamete-like genome packaging process could emerge, following the acquisition of a capsid protein gene from an ancestral RNA virus. Such an event, would allow the reduced cellular genome to be reproduced in many more copies, at the same time relieving the burden of maintaining the viability of the infected host cells. The soma-like virus factory could then become the transient organism we observe today.

In summary, the past years have seen a spectacular renaissance of the field of viral evolution, equally prompted by the publication of increasing bold theories on the origin of life, the realization that viruses are the dominant life form on earth, an exponential increase of genomic data, and the serendipitous discovery of few giant viruses. Viruses have come a long way from being unbecoming to the Tree of Life, to be given a central role in all major evolutionary transitions [5]. The challenge ahead is now to unify the many evolutionary scenarios that have been proposed using hard facts and experimental data, without getting side-tracked by the many spectacular but anecdotal features that individual virus families have incorporated during their long and probably chaotic history.

**Figure Legends.**

**Figure 1.** Iterative nuclear viriogenesis. Starting on the upper left, a primitive DNA virus (a bacteriophage ancestor) get trapped within a RNA cell and become a primitive nucleus. Cellular genes are progressively recruited to the enlarging nucleus due to the advantages of DNA biochemistry. Yet for a while this situation remained instable and reversible, allowing new "pre-eukaryotic viruses" to be created. These viruses eventually re-infect other cells at various stages of this iterative process. Such an hypothetical scheme provides a mechanism for the emergence of various "overlapping" but not monophyletic virus lineages as well as the rapid reassortments of genes from the viral and cellular pools (i.e. before reaching their "Darwinian Threshold" [28])

**Figure 2.** An alternative interpretation of viruses. I propose that the true nature of complex eukaryotic ds-DNA virus (e.g. NCLDV) is found in the transient "virus factory" they produce at each generation rather than in the reproductive particle with which they have been equated. The "virus factory" is proposed to represent the result of the progressive reductive evolution of an obligate parasitic cellular organism, committed to the viral evolutionary pathway by the loss of a functional translation machinery. For the viral organisms, the virus factory exhibits all the properties of the *soma* (left), while the particle state (right) corresponds to the *germen* (*sensu* August Weismann).

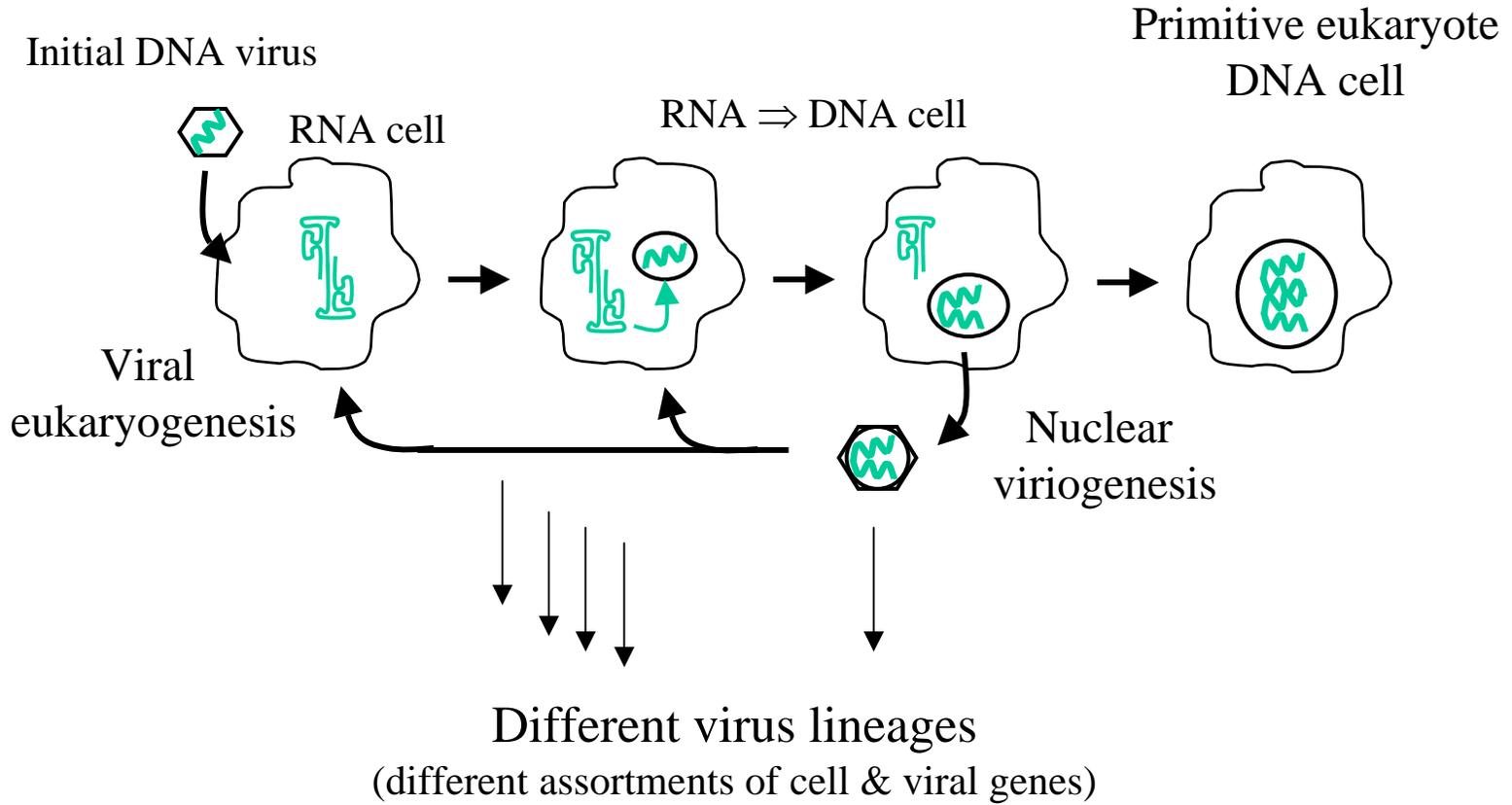

**Fig.1 viral eukaryogenesis and nuclear viriogenesis**

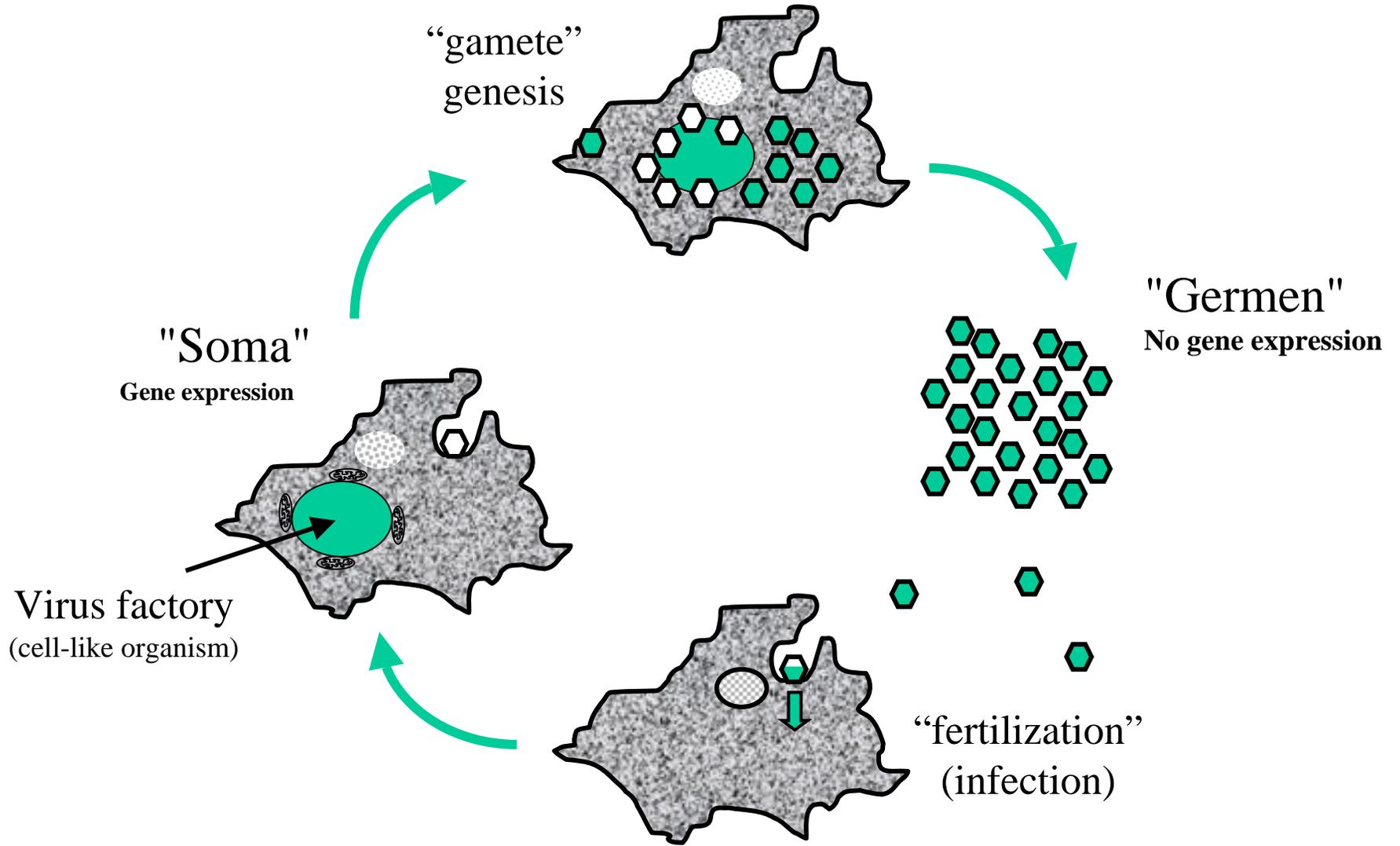

**Fig. 2**